\documentclass[12pt]{article}

\usepackage{color}
\usepackage{graphicx}
\usepackage{amsfonts}
\usepackage{amsthm}
\usepackage{multirow}
\usepackage[top=3cm, bottom=3.25cm, left=2.75cm, right=2.75cm]{geometry}

\newcommand{\Section}[1]{\section{#1} \setcounter{equation}{0}}
\newcommand{\beq}{\begin{equation}}
\newcommand{\eeq}[1]{\label{#1}\end{equation}}
\newcommand{\ber}{\begin{eqnarray}}
\newcommand{\eer}[1]{\label{#1}\end{eqnarray}}
\newcommand{\re}[1]{(\ref{#1})}

\newcommand{\hide}[1]{}

\newcommand{\ft}[2]{{\textstyle\frac{#1}{#2}}}

\newcommand{\ffs}{\mathbb{F}}
\newcommand{\ffbs}{{\bar\mathbb{F}}}
\newcommand{\ffts}{{\tilde\mathbb{F}}}
\newcommand{\fftbs}{{\bar{\tilde\mathbb{F}}}}

\newcommand{\uu}{\mathbb{U}}

\newcommand{\vv}[1]{\mathbb{V}^{#1}}
\newcommand{\vvb}[1]{\bar{\mathbb{V}}^{#1}}
\newcommand{\vvt}[1]{\tilde{\mathbb{V}}^{#1}}
\newcommand{\vvtb}[1]{\bar{\tilde{\mathbb{V}}}^{#1}}

\newcommand{\bbD}[1]{\mathbb{D}_{#1}}
\newcommand{\bbDB}[1]{\bar{\mathbb{D}}_{#1}}
\newcommand{\bbG}[1]{\mathbb{G}_{#1}}
\newcommand{\bbGB}[1]{\bar{\mathbb{G}}_{#1}}
\newcommand{\bbX}[1]{\mathbb{X}_{#1}}
\newcommand{\bbXB}[1]{\bar{\mathbb{X}}_{#1}}

\newcommand{\del}{\delta}
\newcommand{\aleq}{&\!\!\!=\!\!\!&}
\newcommand{\nn}{\nonumber}
\newcommand{\kah}{K\"ahler~}
\newcommand{\pa}[1]{\partial_{#1}}
\newcommand{\lam}{\Lambda}
\newcommand{\G}{\Gamma}
\newcommand{\lamb}{\bar\Lambda}
\newcommand{\lamt}{\tilde\Lambda}
\newcommand{\lamtb}{\bar{\tilde{\Lambda}}}

\newcommand{\eg}{\textit{e.g.},~}

\newcommand{\ep}[1]{\epsilon^{#1}}
\newcommand{\epb}[1]{\bar{\epsilon}^{\,#1}}
\newcommand{\delg}{\delta_g}
\newcommand{\delq}{\delta_Q}
\newcommand{\xip}{\xi^{\+}}
\newcommand{\xim}{\xi^{=}}
\newcommand{\al}{\alpha}
\newcommand{\alb}{\bar\alpha}
\newcommand{\alt}{\tilde\alpha}
\newcommand{\altb}{\bar{\tilde\alpha}}

\newcommand{\+}{{+\!\!\!+}} 
\newcommand{\pp}{\mbox{\tiny${}_{\stackrel\+ =}$}} 

\newcommand{\vp}{\mathbb{V}_{\phi}}
\newcommand{\vc}{\mathbb{V}_{\chi}}
\newcommand{\vpb}{\bar{\mathbb{V}}_{\phi}}
\newcommand{\vcb}{\bar{\mathbb{V}}_{\chi}}

\begin{document}
\renewcommand{\theequation}{\thesection.\arabic{equation}}

\setcounter{page}{0}
\thispagestyle{empty}
\begin{flushright} \small
UUITP-26/10 \\
YITP-SB-10-28\\
\end{flushright}

\smallskip
\begin{center}
 \LARGE
{\bf Off-shell ${ N}=(4,4) $ supersymmetry for new $(2,2)$ vector multiplets}
\\[12mm]
 \normalsize
{\bf M.~G\"oteman${}^a$, U.~Lindstr\"om${}^a$, M.~Ro\v cek${}^b$, and I. Ryb${}^{ab}$} \\[8mm]
 {\small\it
a: Theoretical Physics,\\
Department Physics and Astronomy,\\
Uppsala University, \\ Box 803, SE-751 08 Uppsala, Sweden
\bigskip

b: C.N. Yang Institute for Theoretical Physics,\\
Stony Brook University,\\
Stony Brook, NY 11794-3840, USA}
\end{center}
\vspace{10mm} \centerline{\bfseries Abstract} \bigskip

\noindent We discuss the conditions for extra supersymmetry of the $N=(2,2)$ 
supersymmetric  vector multiplets described in arXiv:0705.3201 [hep-th] and in arXiv:0808.1535 [hep-th]. We find $(4,4)$ supersymmetry for the semichiral vector multiplet but not for the Large Vector Multiplet.
\eject
\tableofcontents

\section{Introduction}
In a recent paper \cite{Goteman:2009ye}, we investigated under what conditions a manifest $N=(2,2)$ sigma model written entirely in terms of left and right semichiral superfields \cite{bbb} admits $N=(4,4)$ supersymmetry \cite{ggg,hklr}.
Here we continue to investigate extended supersymmetry in generalized K\"ahler geometry focusing on the $(2,2)$ vector multiplets introduced in \cite{Lindstrom:2007vc,Lindstrom:2008hx} which were used to gauge isometries and to discuss T-duality in \cite{Lindstrom:2007sq}. See also the related papers \cite{Gates:2007ve} and \cite{Merrell:2007sr}.

We find an off-shell $(4,4)$ algebra for semichiral vector multiplet and present it both at the level of field-strengths and at the level of gauge (pre-)potentials. The algebra closes up to gauge transformations which we calculate. 

For the Large Vector Multiplet (LVM) the situation is different. There cannot exist an off-shell $(4,4)$ supersymmetry for a single LVM. Instead, we find a pseudo-supersymmetry in the abelian LVM, and comment on obstacles in the nonabelian case.

We have organized the paper as follows: In section \ref{multiplets} we recapitulate the basic definitions and properties of the  $N=(2,2)$ vector multiplets. In section \ref{semichiral_section} we present the $N=(4,4)$ supersymmetry for the semichiral vector multiplet and in section \ref{LVM_section} we discuss (twisted) supersymmetry for the Large Vector Multiplet. Conclusions and outlook are contained in the last section.

\Section{Review of the new gauge multiplets}
\label{multiplets}
In this section we review material needed in the rest of the paper.
The most general manifest $N=(2,2)$ supersymmetric sigma model is described by a generalized K\"ahler potential\footnote{The description holds locally and away from irregular points.} which is a function of chiral, twisted chiral and semichiral fields  \cite{Lindstrom:2005zr},
\beq
	K=K(\phi,\bar{\phi},\chi,\bar{\chi}, \bbX{L},\bbXB{L},\bbX{R},\bbXB{R} )~, 
\eeq{potentialK}
where the constraints on the $N=(2,2)$ superfields are
\beq
\bbDB{\pm}\phi=\bbDB{+}\chi = \bbD{-}\chi = \bbDB{+}\bbX{L}=\bbDB{-}\bbX{R}=0~.
\eeq{constraints}
The target space of this $N=(2,2)$ sigma model has been shown \cite{ggg} to admit a bihermitian structure which is equivalent to a generalized K\"ahler structure \cite{Gualtieri:2003dx}.
The simplest isometries act on purely \kah submanifolds of the generalized \kah geometry, 
that is only on the chiral superfields $\phi$ or the twisted chiral superfields $\chi$; 
\beq
K_\phi=K(\phi+\bar{\phi},\dots)~, \quad
K_\chi= K(\chi
+\bar{\chi}, \dots)~.
\eeq{classical_isometry}
These potentials have a rigid isometry,  $\delta\phi=i\lambda$ for $K_\phi$ and $\delta\chi=i\lambda$ for $K_\chi$ with $\lambda$  a real constant parameter. When gauged, however, the local parameter must respect the chirality properties of the superfields,
\beq
\delg \phi = i\lam \quad\Rightarrow\quad
\bbDB{\pm} \lam = 0~;\qquad \delg \chi =i\lam \Rightarrow\quad\bbDB{+} \lam = \bbD{-} \lam = 0~.
\eeq{gauged_isometry}
To ensure the invariance of the Lagrange densities \re{classical_isometry} under the local transformations, we introduce 
the appropriate vector multiplets. These give the well known transformation 
properties for the usual (un)twisted vector multiplets:
\ber
\delta_g V^\phi = i(\lamb-\lam) &\Rightarrow& \delta_g
(\phi+\bar\phi+V^\phi) = 0\nn~,\\
\delta_g V^\chi = i(\lamtb-\lamt) &\Rightarrow&
\delta_g (\chi+\bar\chi+V^\chi) = 0~.
\eer{transf_kahler}
The field-strength is twisted chiral  for $V^\phi$ and chiral for $V^\chi$:
\beq
\tilde W=\bbDB{+}\bbD{-}V^\phi~,\quad W=\bbDB{+}\bbDB{-}V^\chi~.
\eeq{3}
Isometries that involve other combinations of the fields in \re{potentialK} have  been suggested in \cite{Grisaru:1997ep} and recently discussed in \cite{Lindstrom:2007vc, Lindstrom:2007sq}. We now describe the new multiplets that can be used to gauge  these isometries.  

\subsection{The semichiral vector multiplet}
An example of a sigma model with a rigid symmetry that acts on semichiral superfields is given by a potential of the form
\beq
K= K(\phi,\bar{\phi},\chi,
\bar{\chi}, \bbX{L}+\bbXB{L},\bbX{R}+\bbXB{R}, i(\bbX{L}
-\bbXB{L} -\bbX{R}+\bbXB{R}) )~.
\eeq{semichiral_isometry}
The rigid isometry is $\delta \bbX{L}=\delta\bbX{R}=i\lambda$. The gauging of such an isometry requires the chirality properties of the parameters
\beq
 \delta_g \bbX{L} = i \lam_L\ \Rightarrow\ \bbDB{+}
\lam_L = 0~,\quad 
\delta_g \bbX{R} = i \lam_R \ \Rightarrow\ \bbDB{-}
\lam_R = 0 ~.
\eeq{gauged_semi_isometry}
A vector multiplet corresponding to this isometry was introduced in \cite{Lindstrom:2007vc, Gates:2007ve}. This semichiral vector multiplet is
described by three real potentials $(\mathbb{V}^L,\mathbb{V}^R, \mathbb{V}')$ with gauge transformations
\beq
\delg\mathbb{V}^L=i(\bar \Lambda_L-\Lambda_L)~,\quad \delg\mathbb{V}^R=i(\bar \Lambda_R-\Lambda_R)~, \quad \delg\mathbb{V}'=(-\Lambda_L-\bar\Lambda_L+\Lambda_R+\bar\Lambda_R)~,
\eeq{5}
which implies that
\ber\nonumber
&&\delta_g(\bbX{L}+\bbXB{L}+\mathbb{V}^L)=0~,~~ \delta_g(\bbX{R}+\bbXB{R}+\mathbb{V}^R)=0\\[1mm]
&&\delta_g( \bbX{L}-\bbXB{L} -\bbX{R}+\bbXB{R}+i\mathbb{V}')=0~.
\eer{yy}
We construct field-strengths from complex combinations of the real potentials:
\ber
\mathbb{V}_\phi =\ft{1}{2}\left( i\mathbb{V}'+\mathbb{V}^L-\mathbb{V}^R\right)~, && \delg \vp = i(\Lambda_R-\Lambda_L)~,\cr
{\mathbb{V}}_\chi =\ft{1}{2}\left( i\mathbb{V}'+\mathbb{V}^L+\mathbb{V}^R\right)~, && \delg \vc = i(\bar\Lambda_R-\Lambda_L)~.
\eer{7}
These potentials satisfy the reality condition
\beq
\bar \vp- \vp = \bar \vc- \vc~.
\eeq{cndn}
The gauge invariant field-strengths of the semichiral vector multiplet are chiral and twisted chiral superfields,
\beq
\mathbb{F} =i\bbDB{+}\bbDB{-} \mathbb{V}_\phi~,\quad \tilde\mathbb{F} =i\bbDB{+}\bbD{-} \mathbb{V}_\chi~, \quad
\bar{\mathbb{F}} =i\bbD{+}\bbD{-} \bar{\mathbb{V}}_\phi~,\quad \bar{\tilde\mathbb{F}} =i\bbD{+}\bbDB{-} \bar{\mathbb{V}}_\chi~.
\eeq{8}
In the left semichiral representation \cite{Lindstrom:2007vc}, \cite{Lindstrom:2008hx}, we can introduce covariant derivatives as
\ber
 \bar{\nabla}_+ &=& \bbDB{+} \nn \\
 \bar{\nabla}_- \aleq e^{-\vv{}_\phi} \bbDB{-} e^{\vv{}_\phi} \nn \\
  \nabla_+ \aleq e^{-\vv{}_\phi}e^{\vvtb{}_\chi} \bbD{+} e^{-\vvtb{}_\chi} e^{\vv{}_\phi}  
  = e^{-\vvt{}_\chi}e^{-\vvb{}_\phi} \bbD+ e^{\vvb{}_\phi} e^{\vvt{}_\chi}  \nn 
= e^{-\vv L}\bbD+ e^{\vv L}
\\
\nabla_- \aleq e^{-\vvt{}_\chi} \bbD{-} e^{\vvt{}_\chi}~.
\eer{semichiral_der1}
For the abelian case, this gives the following set of connections:
\begin{eqnarray}
\bar{\Gamma}_+ &=& 0 \nn\\
\bar{\Gamma}_- &=& \bbDB{-}\vp \nn\\
\Gamma_- &=& \bbD{-}\vc \nn \\
\Gamma_+ &=& \bbD{+}\vv{L} = \bbD{+}(\vp+\vcb) = \bbD{+}(\vpb+\vc)~.
\label{eq::Gamma_trans}
\end{eqnarray}
Correspondingly, the field-strengths read
\begin{eqnarray}
 \ffs &=& i\{\bar\nabla_+\,,\bar\nabla_-\} =i~\! \bbDB{+} \bar{\Gamma}_- \nn \\
 \ffts \aleq i\{\bar\nabla_+\,,\nabla_-\}=i~\! \bbDB{+} \Gamma_- \nn \\
 \ffbs \aleq -i\{\nabla_+\,,\nabla_-\}=-i~\! (\bbD{+}\Gamma_- + \bbD{-}\Gamma_+) \nn \\
 \fftbs \aleq -i\{\nabla_+\,,\bar\nabla_-\}=-i~\! (\bbD{+}\bar{\Gamma}_- + \bbDB{-}\Gamma_+) ~,
 \label{eq::f_gamma}
 \end{eqnarray}
 where the last equalities again refer to the abelian case.

\subsection{The Large Vector Multiplet}
An example of a sigma model with a rigid symmetry $\delta\phi=\delta\chi=i\lambda$ that acts simultaneously on chiral and twisted chiral superfields is given by a potential of the form
\beq
K=K(\phi+
\bar{\phi},\chi+\bar{\chi} ,i(\phi-\bar\phi-\chi+\bar\chi), 
\bbX{L},\bbXB{L}, \bbX{R},\bbXB{R} )~.
\eeq{LVM_potential}
This isometry is gauged by the Large Vector Multiplet (LVM) with three real potentials $(V^\phi,V^\chi,V')$, where in addition to the gauge transformations of $V^\phi$ and $V^\chi$ in \re{transf_kahler}, the potential $V'$ transforms as
\beq
\delg V'=(-\Lambda-\bar \Lambda+\tilde\Lambda+\bar {\tilde\Lambda})~.
\eeq{4}
As for the semichiral multiplet, we introduce complex combinations
\ber
V_L = \ft{1}{2} (-V^\prime + i (V^\phi-V^\chi)) &\Rightarrow&
\delg V_L = \lam-\lamt~, \nn \\ 
\label{vright} V_R = \ft{1}{2} (-V^\prime + i (V^\phi+V^\chi))~&\Rightarrow&
\delg V_R = \lam-\lamtb~,
\eer{gauge_VL_VR}
subject to the reality condition 
\beq
 V_L + \bar{V}_L =  V_R + \bar{V}_R~.
\eeq{reality_lvm}
We construct gauge invariant field-strengths  for the Large Vector Multiplet as 
\beq
\bbG{+} =\bbDB{+}V_{L}~, \quad \bbG{-} =\bbDB{-}V_{R}~,
\eeq{8b}
and their complex conjugate. The field-strengths for the LVM are thus semichiral spinors. 

\Section{$N\!=\!(4,4)$ susy for the semichiral multiplet}

For simplicity we begin this section with a discussion of the abelian case.
\label{semichiral_section}
\subsection{The abelian semichiral multiplet}
It is well-known that a chiral and a twisted chiral superfield  allow $(4,4)$ supersymmetry \cite{ggg}.\footnote{Various $(4,4)$  models that involve  chiral and twisted chiral superfields   have been discussed in \cite{Gates:1994yk}. The application to the present situation will be described in \cite{GLMR}.} It follows that the   four field-strengths \re{eq::f_gamma} transform under  $(4,4)$ supersymmetry according to \begin{eqnarray}
\delq \ffs &=& \ep{+}\bbDB{+}\fftbs + \ep{-}\bbDB{-}\ffts \nn\\
\delq \ffbs &=& \epb{+}\bbD{+}\ffts + \epb{-}\bbD{-}\fftbs  \nn\\
\delq \ffts &=& -\ep{+}\bbDB{+}\ffbs -\epb{-}\bbD{-}\ffs \nn\\
\delq \fftbs &=& - \epb{+}\bbD{+}\ffs - \ep{-}\bbDB{-}\ffbs~.
\label{eq::F_transformations_new}
\end{eqnarray}
This gives an algebra that closes off-shell;
\begin{equation}
 [\delq (\ep{}_1) , \delq(\ep{}_2)] \ffs = \xip \pa{\+} \ffs + \xim \pa{=} \ffs~,
\end{equation}
where we labeled products of supersymmetry parameters as 
\begin{equation}
\xi^{\pp} = i \ep{\pm}_{[2}\epb{\pm}_{1]}~,
\end{equation}
and the supersymmetry algebra is $\{\bbDB{\pm}, \bbD{\pm}\} = i\pa{}\pp$.
The corresponding transformations on the potentials $\vp$ and $\vc$ can then be deduced using this ansatz together with the constraint on the imaginary part (\ref{cndn}), 
\ber
\delq \vp&=&-\epsilon^- \bbD{-}\vc - \epsilon^+\bbD{+}\bar\vc + \bar\epsilon^- \bbDB{-}\vp - \bar\epsilon^+\bbDB{+}\vp\cr
\delq \vc &=& \epsilon^+\bbD{+}\bar\vp + \bar\epsilon^- \bbDB{-}\vp - \epsilon^- \bbD{-}\vc + \bar\epsilon^+\bbDB{+}\vc~.
\eer{tfsft}
The supersymmetry  algebra closes on the potentials up to gauge transformations according to:
\beq
[\delq(\ep{}_1),\delq(\ep{}_2)]\left(\begin{array}{c}\vp\cr\vc\end{array}\right)
=\left[\xim\pa{=} +\xip\pa{\+}\right]
\left(\begin{array}{c}\vp\cr\vc\end{array}\right)
+\left(\begin{array}{c}\Lambda_R - \Lambda_L
+2\alt\ffts +2\altb\fftbs\cr
\bar\Lambda_R -\Lambda_L + 2\al\ffs + 2\alb\ffbs\end{array}\right)~,
\eeq{close}
where
\ber\nonumber
&&\alpha = i \ep{+}_{[2} \ep{-}_{1]}~,~~~~
\tilde{\alpha}= i \ep{+}_{[2} \epb{-}_{1]}\\[1mm]\nonumber
&&\Lambda_L = -i\xip\, \bbDB{+}\bbD{+}(\bar\vc +\vp)\\[1mm]
&&\Lambda_R = -i\xim\,  \bbDB{-}\bbD{-}(\vc-\vp)~.
\eer{lambd}
Note that the field-strengths appearing in (\ref{close}) have the correct chirality to serve as gauge transformations.

Substituting the transformation laws for the potentials in (\ref{tfsft}) in equation (\ref{eq::Gamma_trans}) 
we find, in the left semichiral representation
\begin{eqnarray}
\delq \bar{\Gamma}_+ &=& 0 \nn \\
\delq \Gamma_+ &=& i~\! \ep{-}\ffbs-i~\! \epb{-}\fftbs   + \ep{-}\bbD{+}\Gamma_- - \epb{-}\bbD{+}\bar{\Gamma}_- \nn\\
\delq \bar{\Gamma}_- &=&  i ~\! \ep{+}\fftbs + i ~\! \epb{+}\ffs +\ep{-} \bbDB{-}\Gamma_-  \nn \\
\delq \Gamma_- &=&    - i ~\! \epb{+}\ffts - i ~\! \ep{+}\ffbs -\epb{-} \bbD{-}\bar{\Gamma}_- ~.
\end{eqnarray}

We go to a real representation using the shift $\delta_g \Gamma = -iD(\Xi)$ where
\beq
\Xi =i(\ep{-}\Gamma_- - \epb{-}\bar\Gamma_-)~.
\eeq{K}
This puts all $\Gamma$ transformations on equal footing:
\begin{eqnarray}
 \delq \bar{\Gamma}_+ &=& i\ep{-}\ffts - i\epb{-}\ffs \nn \\
  \delq \Gamma_+ &=& -i\epb{-}\fftbs + i\ep{-}\ffbs \nn\\
 \delq \bar{\Gamma}_- &=& i\ep{+}\fftbs + i\epb{+}\ffs  \nn \\
 \delq {\Gamma}_- &=& -i\ep{+}\ffbs - i\epb{+}\ffts~. 
 \label{eq::new_Gamma_Transformations}
\end{eqnarray}
It is easy to verify that these transformation laws are 
compatible with the transformation laws for the field strengths (\ref{eq::F_transformations_new}).

\subsection{The nonabelian semichiral vector multiplet}
We now extend the previous discussion to the  nonabelian case \cite{Lindstrom:2007vc}.

For a nonabelian gauge group, the covariant derivatives are given by \re{semichiral_der1} and the field-strengths by the anticommutators in \re{eq::f_gamma}. The nonabelian version of the gauge transformations in \re{5} is
\begin{eqnarray}
 g(\Lambda)\, e^{\vv{L}} &=& e^{i\lamb_L}e^{\vv{L}}e^{-i\lam_L} \nn \\
 g(\Lambda)\,e^{\vv{R}} \aleq e^{i\lamb_R}e^{\vv{R}}e^{-i\lam_R} \nn \\
 g(\Lambda)\,e^{\vp} \aleq e^{i\lam_R}e^{\vp}e^{-i\lam_L} \nn \\
 g(\Lambda)\, e^{\vc} \aleq e^{i\lamb_R}e^{\vc}e^{-i\lam_L} \label{eq::transformations}~.
\end{eqnarray}
The nonabelian extensions of \re{7} and \re{cndn} are
\begin{eqnarray}
e^{\vv{R}} \aleq e^{\vc}e^{-\vp}=e^{-\vpb}e^{\vcb}  \nn \\
e^{\vv{L}} \aleq e^{\vcb}e^{\vp}=e^{\vpb}e^{\vc}~.
\end{eqnarray}

\subsubsection*{Real representation}
\label{realrep}
Finding additional supersymmetries for the semichiral multiplet is facilitated by working in a real representation. Below we present the material needed for this.

We introduce  $\uu_L$ and  $\uu_R$ such that
\ber
 e^{\vv{L}} \aleq e^{\bar\uu_L}e^{\uu_L} ~,\quad g(\Lambda,K)\, e^{\uu_L} 
 = e^{iK}e^{\uu_L}e^{-i\lam_L} \nn \\
 e^{\vv{R}} \aleq e^{\bar\uu_R}e^{\uu_R} 
  ~,\quad g(\Lambda,K)\, e^{\uu_R} = e^{iK}e^{\uu_R}e^{-i\lam_R}~,
\eer{Udef}
with the parameter $K$ an arbitrary Lie algebra valued gauge parameter,  
and thus
\begin{eqnarray}
e^{\vc} &=& e^{\bar\uu_R}e^{\uu_L} \nn \\
e^{\vp} \aleq e^{-\uu_R}e^{\uu_L} ~.
\end{eqnarray}
In the real representation, matter transforms as 
$M \rightarrow e^{iK} M$ and the derivatives read
\begin{eqnarray}
 \bar{\nabla}_+ &=& e^{\uu_L} \bbDB{+} e^{-\uu_L} \nn \\
 \nabla_+ \aleq e^{-\bar\uu_L} \bbD{+} e^{\bar\uu_L} \nn \\
 \bar{\nabla}_- \aleq e^{\uu_R} \bbDB{-} e^{-\uu_R} \nn \\
 \nabla_- \aleq e^{-\bar\uu_R} \bbD{-} e^{\bar\uu_R}~. \label{eq::real_rep_derivatives}
\end{eqnarray}
The gauge transformations are
\begin{eqnarray}
 e^{\uu_L} \delta_g e^{-\uu_L} &=& 
 -iK + e^{\uu_L} i \lam_L e^{-\uu_L} \nn \\
 e^{-\bar\uu_L} \delta_g e^{\bar\uu_L}\aleq 
 -iK + e^{-\bar\uu_L} i \lamb_L e^{\bar\uu_L} \nn \\
 e^{\uu_R} \delta_g e^{-\uu_R} \aleq 
 -iK + e^{\uu_R} i \lam_R e^{-\uu_R} \nn \\
 e^{-\bar\uu_R} \delta_g e^{\bar\uu_R} \aleq 
 -iK + e^{-\bar\uu_R} i \lamb_R e^{\bar\uu_R}~.
\end{eqnarray}
A transformation on $\Gamma$ can be written as
\beq
\delta\Gamma=[\nabla,e^{-V} \delta e^V]~,
\eeq{anyvar}
hence, \eg
\begin{equation}
\delta_g \Gamma_+ =  -i \nabla_+K~.
\end{equation}

In the abelian case,
$ \delta \vv{L} = \delta \uu_L + \delta \bar\uu_L $,
from which we derive the abelian supersymmetry transformations for $\uu_L$,
\begin{equation}
\delq \uu_L = \epb{-}\bbDB{-} (\vp + \vcb)~.
\end{equation}

\subsubsection*{$N=(4,4)$ supersymmetry}
We assume that the supersymmetry transformation laws for the field-strengths 
(\ref{eq::F_transformations_new}) generalize: 
\begin{eqnarray}
 \delq \ffs &=& i (\bar{\nabla}_+ \delq \bar{\Gamma}_- + 
 \bar{\nabla}_- \delq \bar{\Gamma}_+) = 
 \ep{-}\bar{\nabla}_- \ffts + \ep{+}\bar{\nabla}_+ \fftbs \nn \\
 \delq \ffts \aleq i (\bar{\nabla}_+ 
 \delq \Gamma_- + \nabla_- \delq \bar{\Gamma}_+) 
 = -\ep{+}\bar{\nabla}_+ \ffbs - \epb{-}\nabla_- \ffs~.
\end{eqnarray} 
For a covariant derivative of the form 
\beq
\nabla = D +\Gamma =e^{-V} D e^V~,
\eeq{anyd}
an arbitrary variation can be written as in \re{anyvar}
and hence, \eg
\beq
 \del \bar{\G}_+ = [\bar{\nabla}_+,e^{\uu_L}\del(e^{-\uu_L})]~.
\eeq{trans_gamma}
For a supersymmetry transformation, starting from (\ref{eq::new_Gamma_Transformations}),
we thus have
\begin{eqnarray}
 \del_Q \bar{\G}_+ 
 &=& i\ep{-}\ffts - i\epb{-}\ffs \nn \\
 &=& \epb{-}\{ \bar{\nabla}_+, \bar{\nabla}_- \} - 
\ep{-}\{ \bar{\nabla}_+, \nabla_- \} \nn \\
\aleq \epb{-}\{ e^{\uu_L} \bbDB{+} e^{-\uu_L}, \bar{\nabla}_- \} -
 \ep{-}\{e^{\uu_L} \bbDB{+} e^{-\uu_L}, \nabla_- \} \nn \\
\aleq e^{\uu_L} \left( \epb{-}\{ \bbDB{+},e^{-\uu_L}\bar{\nabla}_- e^{\uu_L}\} - 
\ep{-}\{ \bbDB{+},e^{-\uu_L}\nabla_- e^{\uu_L}\}  \right) e^{-\uu_L}\nn \\
\aleq e^{\uu_L} \left( 
\epb{-}\{ \bbDB{+},e^{-\uu_L} [ \bar{\nabla}_- , e^{\uu_L} ] \} - 
\ep{-}\{ \bbDB{+},e^{-\uu_L} [ \nabla_- , e^{\uu_L}] \} 
\right) e^{-\uu_L} \nn \\
\aleq \epb{-}\{ \bar{\nabla}_+,[ \bar{\nabla}_- , e^{\uu_L} ] e^{-\uu_L} \} - 
\ep{-}\{ \bar{\nabla}_+,[ \nabla_- , e^{\uu_L}]  e^{-\uu_L} \}~.
\end{eqnarray}
Identifying this with \re{trans_gamma} we find the supersymmetry transformations for the potentials $\uu_{L,R}$,
\begin{eqnarray}
e^{\uu_L}\del_Q (e^{-\uu_L}) &=& \ep{-}(\nabla_- e^{\uu_L})e^{-\uu_L} -
\epb{-}(\bar{\nabla}_- e^{\uu_L})e^{-\uu_L} \nn \\
e^{\uu_R}\del_Q (e^{-\uu_R}) \aleq \ep{+}(\nabla_+ e^{\uu_R})e^{-\uu_R}-
\epb{+}(\bar{\nabla}_+ e^{\uu_R})e^{-\uu_R} \nn \\
e^{-\bar\uu_L}\del_Q (e^{\bar\uu_L}) \aleq 
\epb{-}(\bar{\nabla}_- e^{-\bar\uu_L})e^{\bar\uu_L} - 
\ep{-}(\nabla_- e^{-\bar\uu_L})e^{\bar\uu_L} \nn \\
e^{-\bar\uu_R}\del_Q (e^{\bar\uu_R}) \aleq 
\epb{+}(\bar{\nabla}_+ e^{-\bar\uu_R})e^{\bar\uu_R} -
\ep{+}(\nabla_+ e^{-\bar\uu_R})e^{\bar\uu_R}  ~,
\end{eqnarray}
which can be rewritten as
\begin{eqnarray}
 (\del_Q e^{-\uu_L})e^{\uu_L} &=&
 \ep{-} e^{-\vc}\bbD{-}e^{\vc} - \epb{-}e^{-\vp}\bbDB{-}e^{\vp} \nn \\
 (\del_Q e^{-\uu_R})e^{\uu_R} \aleq
 \ep{+} e^{-\vcb}\bbD{+}e^{\vcb} - \epb{+}e^{\vp}\bbDB{+}e^{-\vp} \nn \\
 (\del_Q e^{\bar\uu_L})e^{-\bar\uu_L} \aleq
 \epb{-} e^{\vcb}\bbDB{-}e^{-\vcb} - \ep{-}e^{\vpb}\bbD{-}e^{-\vpb} \nn \\
(\del_Q e^{\bar\uu_R})e^{-\bar\uu_R} \aleq
 \epb{+} e^{\vc}\bbDB{+}e^{-\vc} - \ep{+}e^{-\vpb}\bbD{+}e^{\vpb} ~.
\end{eqnarray}
It is now straightforward to calculate the nonabelian supersymmetry transformations for the potentials $\vp$ and $\vc$, 
\begin{eqnarray}
 e^{-\vp} (\del_Q e^{\vp}) 
\aleq e^{-\uu_L}e^{\uu_R} (\del_Q e^{-\uu_R}) e^{\uu_L} + e^{-\uu_L} (\del_Qe^{\uu_L}) \\
\aleq e^{-\vp} (\del_Q e^{-\uu_R})e^{\uu_R} e^{\vp} - (\del_Q e^{-\uu_L})e^{\uu_L} \nn\\
\aleq  \ep{+} e^{-\vp} e^{-\vcb}(\bbD{+}e^{\vcb})e^{\vp} - 
\epb{+} (\bbDB{+}e^{-\vp}) e^{\vp} - 
\ep{-} e^{-\vc}\bbD{-}e^{\vc} + 
\epb{-}e^{-\vp}\bbDB{-}e^{\vp} \nn
\end{eqnarray}
and
\begin{eqnarray}
 e^{-\vc} (\del_Q e^{\vc}) 
\aleq e^{-\uu_L}e^{-\bar \uu_R} (\del_Q e^{\bar \uu_R}) e^{\uu_L} + e^{-\uu_L} (\del_Qe^{\uu_L}) \\
\aleq e^{-\vc} (\del_Q e^{\bar \uu_R})e^{-\bar \uu_R} e^{\vc} - (\del_Q e^{-\uu_L})e^{\uu_L} \nn\\
\aleq  \epb{+}(\bbDB{+}e^{-\vc})e^{\vc} -
\ep{+}e^{-\vc}e^{-\vpb} (\bbD{+}e^{\vpb}) e^{\vc} -
\ep{-} e^{-\vc}\bbD{-}e^{\vc} +
\epb{-}e^{-\vp}\bbDB{-}e^{\vp} .\nn
\end{eqnarray}
\subsubsection*{Closure of the algebra for the field-strengths}
Unlike in the abelian case, where the field-strengths are invariant under gauge transformations, the gauge transformations in the closure of  the supersymmetry algebra here also transform the field-strengths.
Defining
\beq
\{\bar{\nabla}_+ , \nabla_+ \} = i\nabla_{\+}=i\partial_{\+}+i\G_\+ ~,
\eeq{nabla++}
for $\ffs$ we find
\begin{eqnarray}
 [\del_1 ,\del_2]\ffs &=& \del_1 (  \ep{+}_2\bar{\nabla}_+ \fftbs  + \ep{-}_2\bar{\nabla}_- \ffts ) - (1 \leftrightarrow 2) \nn \\
&=& \ep{+}_2 \bar{\nabla}_+ (-\epb+_1 \nabla_+ \ffs  + \ep{-}_1 \bar{\nabla}_- \ffbs)
+ \ep{-}_2 \bar{\nabla}_- (\ep{+}_1 \bar{\nabla}_+ \ffbs - \epb{-}_1 \nabla_- \ffs)  \nn \\  
&& + \ \ep{+}_2(i\ep{-}_1[\ffts,\fftbs]-i\epb{-}_1[\ffs,\fftbs]) + \ep{-}_2 (i\ep{+}_1[\fftbs,\ffts] + i\epb{+}_1[\ffs,\ffts])  \nn \\
&=& \xi^{\+} [\nabla_{\+}, \ffs] +  \xi^{=} [\nabla_{=}, \ffs] + \bar\alpha [\ffbs, \ffs] - \tilde\alpha [\ffts, \ffs] + \bar{\tilde{\alpha}}[\fftbs, \ffs]~.
\end{eqnarray}

Generally, for any $F\in \{ \ffs, \ffbs, \ffts, \fftbs \}$, the supersymmetry algebra closes up to a gauge transformation,
\begin{equation}
 [\del_1,\del_2] F = \xi^{\+} \pa{\+}F + \xi^{=} \pa{=}F 
+ [K(\epsilon_1,\epsilon_2), F]~,
\end{equation}
with the hermitian gauge parameter $K(\epsilon_1,\epsilon_2)$ defined as 
\beq
K(\epsilon_1,\epsilon_2)=\xi^{\+}\G_{\+}+\xi^{=}\G_{=} - \alpha \ffs + \bar{\alpha} \ffbs - \tilde{\alpha} \ffts + \bar{\tilde{\alpha}} \fftbs ~.
\eeq{gaugeparameterK}

The semichiral parameters of eq. (\ref{lambd}) generalize to
\begin{eqnarray}
\Lambda_L&=& -i\xi^{\+} \bar{\nabla}_{+} e^{-\vv{L}}\bbD{+}e^{\vv{L}}~,\cr
\Lambda_R&=&-i\xi^{=} \bar{\nabla}_{-}e^{-\vv{R}}\bbD{-}e^{\vv{R}}~.
\end{eqnarray}

\Section{No $N=(4,4)$ susy for the Large Vector Multiplet}
\label{LVM_section}
We now turn to the Large Vector Multiplet which gauges isometries acting simultaneously on  chiral and twisted chiral superfields, as described in section \ref{multiplets}. As we shall see, the situation is completely different as compared to the semichiral multiplet discussed in the previous section.
For the Large Vector Multiplet (LVM), the field-strengths are semichiral fermionic superfields. This fact implies that there is no $N=(4,4)$ off-shell supersymmetry in the abelian case \cite{Goteman:2009ye, Goteman:2009xb}. 

The field-strengths for the abelian LVM defined in \re{8b} are four semichiral fermions,
\beq
\bbG{+}=\bbDB{+}V_{L}~, \quad \bbGB{+}=\bbD{-}\bar{V}_{L}~,\quad \bbG{-}=\bbDB{-}V_{R}~, \quad \bbGB{-}=\bbD{-}\bar{V}_{R}~.
\eeq{8b_again}
As shown in paper \cite{Goteman:2009xb} and discussed further in \cite{Goteman:2009ye}, additional off-shell $N=(4,4)$ supersymmetry can only be imposed on a set of semichiral fields if the dimension of the space of fields is larger than four. It is therefore impossible to impose additional supersymmetry on the field-strengths of the abelian Large Vector Multiplet.

\subsubsection*{Left/right supersymmetry}
Even though we cannot impose ordinary $N=(4,4)$ supersymmetry, we could make an ansatz for independent \emph{left} susy for $\bbG{+}$ and \emph{right} susy for $\bbG{-}$ as
\begin{eqnarray}
\delq \bbG{+} &=& \epb{-}\bbDB{-}\bbG{+} - \ep{-}\bbD{-}\bbG{+}\nn \\
\delq \bbG{-} &=& \epb{+}\bbDB{+}\bbG{-} - \ep{+}\bbD{+}\bbG{-}~,
\label{eq::G_transformation_half}
\end{eqnarray}
which closes to 
\beq
[\delq(\ep{}_1),\delq(\ep{}_2)]\bbG{+} = \xi^{=} \pa{=}\bbG{+}~, \quad [\delq(\ep{}_1),\delq(\ep{}_2)]\bbG{-} = \xi^{\+} \pa{\+}\bbG{-}~.
\eeq{Gclosure_abelian_half}

\subsubsection*{Twisted supersymmetry}
Another option is to impose additional \emph{pseudo}-supersymmetry. An ansatz for the semichiral field-strengths of the Large Vector Multiplet is
\begin{eqnarray}
 \delta \bbG{+} &=& \epb{-}\bbDB{-}\bbG{+} +  \ep{-}\bbD{-}\bbG{+} + \ep{+}\bbDB{+}\bbGB{+}\nn \\
 \delta \bbGB{+} &=& \ep{-}\bbD{-}\bbGB{+} + \epb{-}\bbDB{-}\bbGB{+} + \epb{+}\bbD{+}\bbG{+}\nn \\
 \delta \bbG{-} &=& \epb{+}\bbDB{+}\bbG{-} + \ep{+}\bbD{+}\bbG{-} + \ep{-}\bbDB{-}\bbGB{-} \nn \\
\delta \bbGB{-} &=& \ep{+}\bbD{+}\bbGB{-} + \epb{+}\bbDB{+}\bbGB{-} + \epb{-}\bbD{-}\bbG{-}~.
\label{eq::G_transformation_twisted}
\end{eqnarray}
This ansatz closes to a pseudo-supersymmetry algebra
\beq
 [\delta(\ep{}_1),\delta(\ep{}_2)]\bbG{+} = -\bigr(\xi^{\+} \pa{\+}\bbG{+} + \xi^{=} \pa{=}\bbG{+}\bigr).
 \eeq{Gclosure_abelian_twisted}
Identifying the field-strength potential $\bbG{+} = \bbDB{+}V_L$ and $\bbG{-} = \bbDB{-}V_R$ and using the reality constraint $V_L + \bar{V}_L = V_R + \bar V_R$, we find
\begin{eqnarray}
\delta V_L &=&  \epb{-} \bbDB{-} V_L + \ep{-} \bbD{-} V_L - \ep{+}\bbD{+}\bar{V}_L + \epb{+}\bbDB{+}(2V_L + \bar{V}_L)~,\nn \\
\delta V_R &=& \epb{+} \bbDB{+} V_R + \ep{+} \bbD{+} V_R - \ep{-}\bbD{-}\bar{V}_R + \epb{-}\bbDB{-}(2V_R + \bar{V}_R)~. 
\end{eqnarray}
The transformations close to a pseudo-supersymmetry algebra.

In \cite{Goteman:2009xb}, we show that $N=(4,4)$ supersymmetry can be imposed for a set of semichiral fields if their number is $4d$ with $d>1$. This might suggest that it is possible to find extra supersymmetries for the nonabelian Large Vector Multiplet. However, for a gauge multiplet  the extra supersymmetry should commute with the gauge transformations. We do not believe that this is possible, at least not for an off-shell supersymmetry algebra. There are reasons from T-duality to believe that it might be possible to find an on-shell $(4,4)$-algebra for certain actions, but we have not investigated that option in detail.

\Section{Conclusions}
\label{conclusions_section}
In this paper, we discuss $(4,4)$ supersymmetry for two gauge multiplets introduced in \cite{Lindstrom:2007vc} and \cite{Lindstrom:2008hx}; the semichiral multiplet and the Large Vector Multiplet. 

For the semichiral vector multiplet, we  find off-shell  $(4,4)$ supersymmetry, first for a $U(1)$ and then for a general gauge group. The transformations for the potentials are deduced from those of the field-strengths and close to a supersymmetry up to gauge transformations. 
In \cite{TartaglinoMazzucchelli:2009jn} a related multiplet was recently constructed in bi-projective superspace \cite{Lindstrom:1994mw}. We believe that a suitable partial gauge-fixing relates the two, but have not checked this. 

For the Large Vector Multiplet we do not find off-shell $(4,4)$ supersymmetry but are able to construct twisted $(4,4)$ supersymmetry in the abelian case. 
This is in agreement with the results in \cite{Goteman:2009ye, Goteman:2009xb}, where it was shown that $(4,4)$ supersymmetry can be imposed on a set of semichiral fields parametrizing a $d$-dimensional target space  only if $d>4$. 

\bigskip

{\bf Acknowledgements:}\\
The authors wish to thank Warren Siegel for discussions. IR  is happy to thank Uppsala University for hospitality. The research of UL was supported by VR grant 621-2009-4066 and that of MR was supported in part by NSF grant no. PHY-06-53342.
The stimulating atmosphere at the 2010 Simon's summer workshop helped us to finish this paper.
\appendix

\section{Hermiticity conventions}

We are using the conventions of \cite{Gates:1983nr}. Briefly, this implies that a spinor with an upper index is hermitian as is a covariant derivative with a lower index:
\beq
(\psi^\pm)^\dagger=\bar\psi^\pm~, ~~(\bbD{\pm})^\dagger=\bbDB{\pm}~.
\eeq{h}
This implies that  for an operator,
\beq
(\epsilon^\pm\bbD{\pm})^\dagger=-\bar\epsilon^\pm\bbDB{\pm}~,
\eeq{he}
which mimics the relation  for a bosonic derivative and parameter $(\xi^a\partial_a)^\dagger=-\bar\xi^a\partial_a$.
The latter is the familiar fact that as an operator the bosonic derivative $\partial_a$ is antihermitian. Nevertheless, of course $\partial_a f$ is real for a real function $f$. We shall find a similar situation in the fermionic case.

In \cite{Gates:1983nr} the infinitesimal transformation of a superfield $\phi$ is defined via a commutator,
\beq
\delta\phi =[\epsilon^\pm\bbD{\pm},\phi ]~.
\eeq{her}
Using the relation \re{h} this implies that 
\beq
\delta\bar\phi=(\delta\phi)^\dagger =[\epsilon^\pm\bbD{\pm},\phi ]^\dagger =[\bar\epsilon^\pm\bbDB{\pm},\bar\phi ]~.
\eeq{herm}
Assuming that $\epsilon^\pm\bbD{\pm}$ acts on $\psi$ via a commutator, the relations \re{her} and  \re{herm} are written
\beq
\delta\phi=\epsilon^\pm\bbD{\pm}\phi~,\quad \delta\bar\phi=\bar\epsilon^\pm\bbDB{\pm}\bar\phi~.
\eeq{hermi}

There is another way of arriving at the expressions in \re{hermi} that avoids explicitly considering the adjoint of operators. Using 
\beq
D_\theta \theta =1~,
\eeq{hermii}
and considering a constant anticommutig spinor $\eta$, we have the relation
\beq
\epsilon D_\theta\theta\eta = \epsilon\eta~.
\eeq{hermit}
 Conjugating this and remembering that $(\epsilon\eta)^\dagger=\bar\eta\bar\epsilon=-\bar\epsilon\bar\eta$  
we find
\beq
(\epsilon D_\theta)^\dagger(-\bar\theta\bar\eta)=-\bar\epsilon\bar\eta~,
\eeq{hermiti}
from which it follows that $(\epsilon D_\theta\phi)^\dagger=\bar\epsilon \bar D_\theta\bar\phi$  for a superfield $\phi$, and we recover \re{hermi} without involving  commutators of operators.

\end{document}